\documentclass{article}
\usepackage{spconf,amsmath,graphicx}
\usepackage{algorithm}
\usepackage{algorithmic}
\usepackage{booktabs}

\title{Optimizing Dysarthria Wake-Up Word Spotting: An End-to-End Approach for SLT 2024 LRDWWS Challenge}
\name{Shuiyun Liu$^{1}$, Yuxiang Kong$^{2}$, Pengcheng Guo$^{1}$, Weiji Zhuang$^2$, Peng Gao$^2$, Yujun Wang$^2$, Lei Xie$^{1*}$
\thanks{* Corresponding author.}}
\address{$^1$Audio, Speech and Language Processing Group (ASLP@NPU), School of Computer Science,\\
Northwestern Polytechnical University, Xi'an, China\\
$^2$ Xiaomi Inc., China}
\begin{document}
%
\maketitle
\begin{abstract}
Speech has emerged as a widely embraced user interface across diverse applications.
However, for individuals with dysarthria, the inherent variability in their speech poses significant challenges.
This paper presents an end-to-end Pretrain-based Dual-filter Dysarthria Wake-up word Spotting (PD-DWS) system for the SLT 2024 Low-Resource Dysarthria Wake-Up Word Spotting Challenge. 
Specifically, our system improves performance from two key perspectives: audio modeling and dual-filter strategy.
For audio modeling, we propose an innovative 2branch-d2v2 model based on the pre-trained data2vec2 (d2v2), which can simultaneously model automatic speech recognition (ASR) and wake-up word spotting (WWS) tasks through a unified multi-task finetuning paradigm.
Additionally, a dual-filter strategy is introduced to reduce the false accept rate (FAR) while maintaining the same false reject rate (FRR).
Experimental results demonstrate that our PD-DWS system achieves an FAR of 0.00321 and an FRR of 0.005, with a total score of 0.00821 on the test-B eval set, securing first place in the challenge.
\end{abstract}
\section{Introduction}
\label{sec:intro}

Major advances in voice technology have revolutionized human-computer interaction. This technology facilitates hands-free operation, improves accessibility, and enhances the user experience by allowing users to issue commands, control applications, and manage devices through simple voice interaction. Keyword spotting (KWS), particularly wake-up word spotting, is crucial as the initial step in voice interaction~\cite{DBLP:conf/icassp/GaoMSMV20, DBLP:conf/icassp/SahaiLMSASMRCMK23}. In the development of wake-up word spotting technology, the integration of deep learning algorithms significantly improves the accuracy and efficiency of recognition. These algorithms can accurately identify wake-up words in noisy environments and across various accents, thereby providing a more natural and seamless user experience~\cite{DBLP:conf/icassp/CasebeerWS24,DBLP:conf/interspeech/00010L0Y23}.

However, the development of speech recognition technology and even wake-up word spotting (WWS) technology presents potential difficulties for patients. Dysarthria is a motor speech disorder, typically caused by neurological conditions that impair the control of speech muscles and is commonly seen in conditions such as Parkinson's disease, cerebral palsy, and amyotrophic lateral sclerosis (ALS). 
Individuals with dysarthria often exhibit inaccurate articulation, irregular speech rate, disrupted speech rhythm, and decreased volume and clarity~\cite{DBLP:conf/interspeech/YeeLNHTBF23}. 
While devices sold on the market use sophisticated speech recognition technology, they are primarily designed for users with mostly standard, intelligible speech.
This results in a lack of sensitivity to non-standard or unclear speech, significantly degrading recognition performance~\cite{DBLP:journals/jrie/RussisC19}. 
By recognizing dysarthric speech, the communication and interaction abilities of people with this disorder can be significantly enhanced, thereby improving their overall quality of life. Consequently, Dysarthric Speech Recognition (DSR) has garnered considerable attention and interest from researchers worldwide~\cite{DBLP:conf/icassp/YueLCCB22,DBLP:conf/iscslp/WangYWSLM21,DBLP:conf/icassp/HuXJGWCDLM23,DBLP:journals/taslp/LiuGHXCYLM21}. 
\begin{figure*}[ht]
\vspace{-1pt}
\begin{center}
\includegraphics[width=1.0\textwidth]{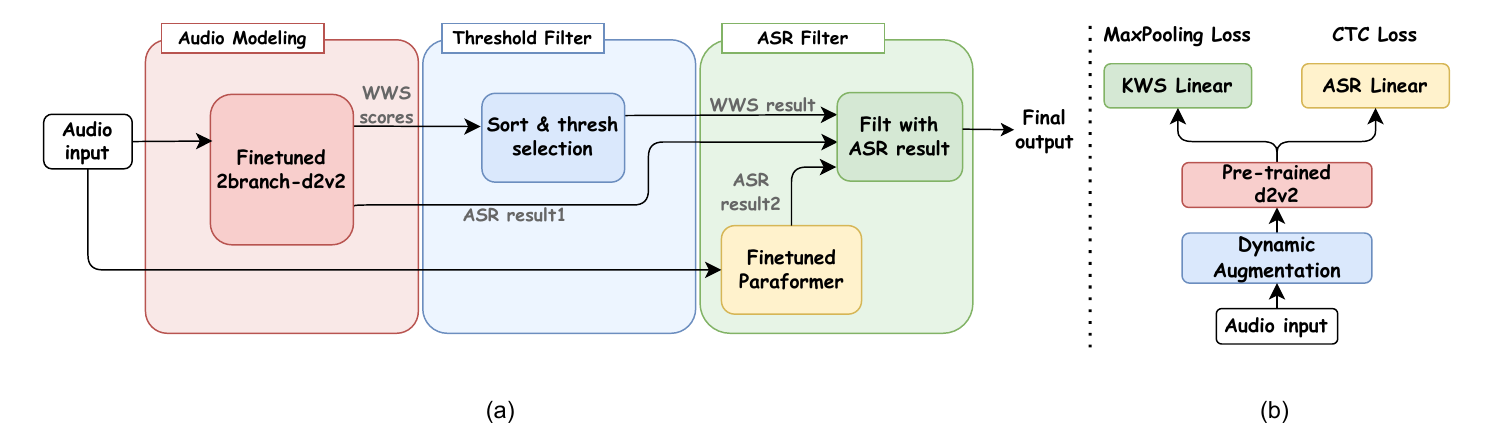}
\end{center}
\vspace{-0.4cm}
\caption{(a) An overview of our proposed PD-DWS system; (b) Details of the 2branch-d2v2 encoder.} 
\label{fig:pdwws}
\vspace{-0.4cm}
\end{figure*}

Early studies mainly focused on the phonological repair of dysarthria.
Yang et al.~\cite{DBLP:conf/biostec/YangC20} employ a cycle GAN network to transform the original dysarthric speech signals in the spectral domain and synthesize new speech signals from the training model, which improves the intelligibility of the language. Daniel et al.~\cite{DBLP:conf/interspeech/KorzekwaBKDL19} utilize a Variable Auto-Encoder (VAE) to reconstruct ambiguous speech signals, which enhances recognition accuracy. However, as the severity of the dysarthria increases, the efficacy of these repair methods diminishes.
It has been shown that incorporating dysarthria data during training, especially personalized models trained using end-user speaker samples, can effectively improve the accuracy of personalized dysarthria models even in severely dysarthric environments~\cite{DBLP:conf/interspeech/GreenMJCHCSLTBN21,DBLP:conf/interspeech/ShorELTBCVMCNHM19,DBLP:conf/icassp/DoshiCJZBRCRM21}. Nonetheless, the scarcity of training data for dysarthria exacerbates the challenge of DSR~\cite{DBLP:conf/interspeech/WongYCWLM15}. To mitigate the data scarcity, some studies have improved recognition performance by leveraging synthetic data for data augmentation~\cite{DBLP:journals/corr/abs-2308-08438,DBLP:journals/corr/abs-2201-11571}, and some studies have utilized the Wav2Vec~\cite{DBLP:conf/nips/BaevskiZMA20} self-supervised speech representations as features for training~\cite{DBLP:conf/interspeech/HernandezPNOMY22}.
Despite these efforts, DSR still faces significant challenges. More regrettably, there is even less research and data on the recognition of dysarthric wake-up words.

 To address this problem, the IEEE SLT 2024 workshop launched the Low Resource Dysarthric Wake Word Recognition (LRDWWS) Challenge~\cite{gao2024enhancingvoicewakeupdysarthria}. The challenge aims to solve speaker-dependent wake-up word spotting tasks using a small amount of wake-up word audio from a specific person. Not only does this research have the potential to improve the quality of life for people with dysarthria, but it could also help smart devices to better meet the needs of different users, making it a truly universal technology. To support this effort, they present the first speech dataset consisting of dysarthric wake-up words in Mandarin in the Challenge and hope to use it to customize the best performing wake-up word spotting system for people with dysarthria.

This study details our participation in the LRDWWS challenge, focusing on the development of a dysarthric wake-up word system named Pretrain-based Dual-filter Dysarthria Wake-up word Spotting (PD-DWS).
Our efforts encompass two key areas: audio modeling and a dual-filter strategy.
Firstly, in the audio modeling part, we introduce an innovative 2branch-d2v2 model by finetuning the pre-trained data2vec2 (d2v2) model within a multi-task framework, which simultaneously models both automatic speech recognition (ASR) and wake-up word spotting (WWS) tasks.
Subsequently, a dual-filter module is proposed to process the model outputs.
Specifically, the output from the WWS branch is sent to the threshold filter, while the ASR branch output is directed to the ASR filter for further refinement.
The threshold filter performs initial filtering on the wake-up word probabilities, preliminarily determining the audio's predicted label. 
The ASR filter then conducts secondary filtering using the ASR output from the model as well as ASR results obtained from the finetuned Paraformer~\cite{DBLP:conf/interspeech/GaoZ0Y22}.
In addition, we finetune the Paraformer with TTS synthesized dysarthric audio, which allows the model to be more adaptable to the dysarthric environment.
By integrating these strategies, our proposed PD-DWS achieves a false accept rate (FAR) of 0.00321 and a false reject rate (FRR) of 0.005, with total scores of 0.00821 on the test-B eval set in this Challenge.
The main contributions of our work are outlined as follows:
\begin{itemize}
  \setlength{\itemsep}{4pt}
  \setlength{\parskip}{-3pt}
    \item We validate the audio modeling capabilities of different encoders, and experiments show that good performance can be achieved in low-resource scenarios using pre-trained models.
    \item The proposed 2Branch-D2V2 model trains both ASR and WWS. And our system employs a two-level filtering mechanism to effectively ensure a low FAR.
    \item Our system utilizes TTS generation to generate corresponding audio for the Finetuned Paraformer module.
    \item The results of the experiment show that our PD-DWS system in the track wins the first place.
\end{itemize}

\section{PROPOSED SYSTEM}
\label{sec:format}

Fig.~\ref{fig:pdwws} (a) overviews our proposed PD-DWS system which comprises an audio modeling and a dual-filter, which includes a threshold filter and an ASR filter.

\subsection{Audio Modeling}
\label{ssec:audioenc}
In the audio modeling part, we explore two different encoder architectures: the Conformer~\cite{gulati2020conformer} encoder and a novel 2branch-d2v2 encoder. The Conformer adds a convolutional module to the self-attention mechanism~\cite{DBLP:conf/nips/VaswaniSPUJGKP17}, allowing both global and local modeling capabilities to be exploited, and achieves better results in different ASR tasks.

This section primarily focuses on introduction and implementation of the 2branch-d2v2 approach.
As shown in Fig.~\ref{fig:pdwws} (b), the 2branch-d2v2 encoder is initialized with a pre-trained d2v2 model~\cite{DBLP:conf/naacl/OttEBFGNGA19}.
This pre-trained model is then finetuned within a multi-task learning framework to optimize its performance for our specific application.

The finetuning process involves directing the output of the d2v2 model into two distinct branches. One branch is dedicated to ASR, while the other is dedicated to WWS. Each branch is trained with a different loss function tailored to its task.
The WWS branch uses the official max pooling loss~\cite{DBLP:conf/icassp/HouSOH020}, denoted as $\mathcal{L}_{\text{WWS}}$, to effectively capture and optimize wake-up word spotting. On the other hand, the ASR branch utilizes the Connectionist Temporal Classification (CTC) loss~\cite{DBLP:conf/icml/GravesFGS06}, denoted as $\mathcal{L}_{\text{CTC}}$, to improve speech recognition accuracy.
The final training loss for the 2branch-d2v2 encoder is a combination of these two loss functions. This composite loss ensures that the model is optimized concurrently for both wake word spotting and speech recognition tasks, leveraging ASR to assist WWS modeling. The formula for the final training loss is:
\begin{equation}
L = 0.5 \cdot L_{\text{CTC}} + 1.0 \cdot L_{\text{WWS}}
\end{equation}

In addition to the core components of our system, we employ dynamic augmentation techniques to enhance model robustness. These techniques include a 10\% variation in audio volume, 15\% dynamic noise addition from the MUSAN~\cite{DBLP:journals/corr/SnyderCP15} dataset with a Signal-to-Noise Ratio (SNR) range of 8 to 20 decibels, and speed perturbation with playback rates varying between 0.9 and 1.1 times the original speed. These augmentations improve the model's resilience to different loudness levels, background noises, and speaking speeds, ensuring reliable performance in diverse acoustic conditions.

We follow the official training data flow\footnote{https://github.com/greeeenmouth/LRDWWS}. In the first step, we train a speaker-independent control (SIC) KWS model from scratch using the control (non-dysarthric) dataset. In the second step, we finetune the SIC model with uncontrol (dysarthric) dataset to obtain a speaker-independent dysarthric (SID) KWS model~\cite{9404373}.
But in the third step, we finetune the SID model using all enrollment sets instead of using separate sets for each individual.

\subsection{Dual-Filter: Threshold Filter}
\label{ssec:selector1}
The threshold filter module operates by processing the probabilities assigned to ten wake-up words, which it receives from the Wake Word Spotting (WWS) branch. For each audio sample, the module receives probabilities for these ten wake-up words. It selects the highest probability among them, designating this maximum probability as the temporal score for the audio and assigning the corresponding wake-up word as the temporal label.

Once the temporal scores and labels are determined, the audio samples are organized in descending order according to their temporal scores. The module then examines these ordered scores to establish provisional thresholds. Specifically, for each audio sample with a temporal score below the determined threshold, the module changes the temporal label to a filter label. Audio samples with scores above the threshold retain their original temporal labels.

Extensive experiments were conducted using the test-A evaluation set to determine the optimal threshold. These experiments reveal that setting the threshold at the 60th highest score among the sorted temporal scores yields the best performance.

\subsection{Dual-Filter: ASR Filter}
\label{ssec:selector2}

The ASR filter module is designed to correct the WWS results from the preceding step by utilizing ASR outputs. 
This module employs two distinct of ASR results for comparison. 
The first set of ASR results is obtained through beam search decoding of the ASR branch within our model. 
The second set is derived from the ASR results produced by the open-source Paraformer large model\footnote{https://github.com/modelscope/FunASR}, which has been finetuned using competition data and TTS-synthetic speech.
The process of correcting the WWS results begins with a comparison of the lengths of detected wake-up words against the lengths of the ASR results. If the length of a wake-up word matches the length of any ASR result, this wake-up word result is retained. Conversely, if there is no match in length, the wake-up word result is discarded and labeled as a filler. 
The detailed methodology for this approach is outlined in Algorithm~\ref{alg1}.

\begin{algorithm}

\caption{Re-correct WWS results with ASR results}
\label{alg1}
\begin{algorithmic}
   \REQUIRE {stage2 predicted word label $p$}
   \REQUIRE {ASR\_result1  , ASR\_result2}
   \STATE wake\_words\_list $\gets$ ten wake-up words 
   \IF {p in wake\_words\_list}
   \IF{len(ASR\_result1) = len($p$) \textbf{or} len(ASR\_result2) = len($p$)}
   \STATE $predicted\_updated \gets p$
   \ELSIF{len(ASR\_result1) $\neq$ len($p$) \textbf{and}  len(ASR\_result2) $\neq$ len($p$)}
   \STATE $predicted\_updated \gets -1$ 
   \ENDIF
   \ENDIF
\end{algorithmic}
\end{algorithm}
\subsection{TTS Generator}
\label{ssec:tts}
\begin{figure}[ht]
\vspace{-1pt}
\begin{center}
\includegraphics[width=0.42\textwidth]{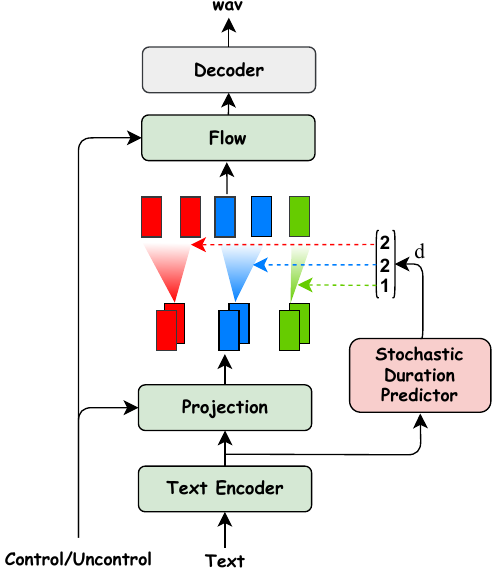}
\end{center}
\vspace{-0.4cm}
\caption{The VITS~\cite{DBLP:conf/icml/KimKS21} system diagram inference procedure. } 
\label{fig:vit}
\vspace{-0.4cm}
\end{figure}

During the finetuning phase of Paraformer, we use TTS data for data augmentation. First, we utilize both the control dataset and the uncontrol dataset to train an end-to-end VITS~\cite{DBLP:conf/icml/KimKS21} model. Specifically, we use control and uncontrol labels to differentiate between various speech styles and incorporate these labels as style embeddings into the text encoder and flow of the VITS model. Using uncontrol label in the inference process can generate audio with dysarthria. Fig.~\ref{fig:vit} shows the details of our inference procedure.

\section{EXPERIMENT CONFIGURATION}
\label{sec:exp}
\subsection{Datasets}
Except for the pre-training phase of the d2v2 model, where additional datasets are used, all other training stages utilize only the LRDWWS training set. For evaluation, we use the LRDWWS eval set. 

The LRDWWS dataset comprises 18,630 recordings totaling 17 hours. This includes 10,125 recordings from non-dysarthric speakers (control), amounting to 7.6 hours, and 8,505 recordings from dysarthric speakers (dysarthria), totaling 9.4 hours. The dataset features speech from 21 dysarthric speakers (12 female, 9 male) and 25 non-dysarthric speakers (13 female, 12 male). 
Each speaker contributes 405 recordings, including 50 wake-up word recordings (10 different wake-up words, with 5 recordings per word) and 355 non-wake-up word recordings. The non-wake-up word recordings consist of fixed command words, descriptions of furniture, TV and audio control words, numbers, interactions, and negative samples. Table~\ref{ress} provides detailed information about the datasets used in training each module.

\begin{table}[t]
\centering
\caption{Details of the training data used for each module.}
\vspace{5pt}
\renewcommand{\tabcolsep}{0.2cm}
\renewcommand\arraystretch{1.3}
\begin{tabular}{ll}
\toprule     
\textbf{Module}                                                              & \textbf{Training data}  \\ 
\midrule 
pretrained\_d2v2                                                     & \begin{tabular}[c]{@{}l@{}}LibriHeavy~\cite{DBLP:journals/corr/abs-2309-08105}, \\GigaSpeech~\cite{DBLP:conf/interspeech/ChenCWDZWSPTZJK21}, \\WenetSpeech~\cite{DBLP:conf/icassp/ZhangLGSYXXBCZW22},\\ Aishell\{1,2\} \cite{DBLP:conf/ococosda/BuDNWZ17,DBLP:journals/corr/abs-1808-10583}, \\
ACAV100M~\cite{DBLP:conf/iccv/LeeCYKBCS21},\\OpenSLR \{38,47,68,82,87,111,\\ 119,123,124,133\}, \\CommonVoice,\\ LRDWWS training set\end{tabular} \\ \midrule
finetuned\_2branch-d2v2                                              & \begin{tabular}[c]{@{}l@{}}LRDWWS training set\\ LRDWWS enrollment set\end{tabular}  \\ \midrule
\begin{tabular}[c]{@{}l@{}}TTS-generator\\ 
\end{tabular} & LRDWWS training set \\ \bottomrule    
\end{tabular}
\label{ress}
\end{table}

\label{ssec:data}
\subsection{Configuration}

\label{ssec:config}
The model configurations used in the experiments are as follows. 
For the baseline model, we re-implement the official base model provided in the challenge. The baseline is based on the WEKWS~\cite{DBLP:conf/icassp/WangXHZZXP23} toolkit framework, using 80-dimensional log Mel-filter banks with a 25ms window and a 10ms shift for input audio signals. In the encoder module, we use a four-layer Depthwise Separable Temporal Convolutional Network (DS-TCN)~\cite{DBLP:conf/icassp/CouckeCGLPL19} with a hidden dimension of 256. We use the Adam optimizer, and each of the three datasets (control, uncontrol, and enrollment) is iterated for 80 epochs.

For our proposed audio modeling part, the conformer encoder has 12 layers, 4 attention heads, 256 hidden dimensions, and approximately 31M parameters.
For the d2v2 pre-training, we utilize the same configuration as the d2v2 large model\footnote{https://github.com/facebookresearch/fairseq}, with a prenet depth of 8, a main part depth of 16, 16 heads, and a dimension of 1024, totaling approximately 300M parameters. Additionally, we set the gradient accumulation steps to 6 and train on 8 A800 GPUs for 600,000 steps. The learning rate scheduler is cosine\_lr, reaching a peak of 0.0004 after 10,000 steps.

For our proposed 2branch-d2v2 finetuning, the model is trained with a dynamic batch size on 2 A800 GPUs, with each batch lasting approximately 300 seconds. We use the Adam optimizer for finetuning. The learning rate for the non-pre-trained parts reaches a maximum of 0.001 after 450 steps, while the learning rate for the pre-trained parts reaches a maximum of 0.00005 after 1600 steps. Each finetuning session continues until convergence, with accuracy (as provided by the official method) reaching approximately 1.0. The three datasets (control, uncontrol, and enrollment) are iterated over approximately 35, 7, and 7 epochs, respectively.

The vocabulary used in the ASR branch is derived from the text in the LRDWWS training set, incorporating characters for Chinese and letters for English, totaling 451 units, including \textless blank\textgreater, \textless unk\textgreater, \textless sos/eos\textgreater.

\subsection{Evaluation}
We use the same metric as the official challenge, evaluating all systems based on the combination of FRR and FAR~\cite{DBLP:conf/icassp/WangCFL23}. This metric mitigates the possibility of overly optimistic evaluations stemming from highly imbalanced class distributions and is defined as follows:
\begin{equation}
\text{Score} = \text{FRR} + \text{FAR} = \frac{N_{\text{FR}}}{N_{\text{wake}}} + \frac{N_{\text{FA}}}{N_{\text{non-wake}}}
\end{equation}

In this evaluation, $N_{\text{wake}}$ represents the number of samples that contain wake-up words, while $N_{\text{non-wake}}$ represents the number of samples without wake-up words. $N_{\text{FR}}$ indicates the number of samples that have a wake-up word but are not recognized as such by the system. Conversely, $N_{\text{FA}}$ represents the number of samples that do not contain wake-up words but are incorrectly identified as positive by the system.

\label{ssec:eval}

\section{RESULTS AND ANALYSIS}
\subsection{Comparison with different base model}
\label{ssec:res}
We conduct experiments on baseline models using the test-A-eval set, as depicted in Table~\ref{sb:exp1}.
The baseline model achieves a score of 0.3112. Our proposed system, whether employing the conformer or 2branch-d2v2, consistently outperforms the baseline model.
Specifically, 2branch-d2v2 achieves a FAR of 0.0043 and an FRR of 0.0300, demonstrating superior performance compared to the conformer, which records a FAR of 0.0183 and an FRR of 0.0825.
Therefore, based on these results from the base model selection section, subsequent experiments will focus on 2branch-d2v2.

\begin{table}[t]
\centering
\caption{Performance of different systems on the test-A-eval set using exhaustive threshold searching.}
\vspace{5pt}
\renewcommand{\tabcolsep}{0.35cm}
\renewcommand\arraystretch{1.3}
\begin{tabular}{cccc}
\toprule     
\textbf{Base model}           & \textbf{Score $\downarrow$}            & \textbf{FAR $\downarrow$}               & \textbf{FRR $\downarrow$} \\ 
\midrule 
Baseline                                         & 0.3112          & 0.0387          & 0.2725 \\
Conformer                                      & 0.1008          & 0.0183 & 0.0825 \\
2branch-d2v2                                     & \textbf{0.0343}          & \textbf{0.0043}          & \textbf{0.0300} \\
\bottomrule    
\end{tabular}
\label{sb:exp1}
\end{table}

\subsection{Comparison with other competition systems}
\label{ssec:othher}
Table \ref{sb:expcop} presents the score results of the official baseline and each competition system. Our system achieves a FAR of 0.003210, an FRR of 0.005000, and a score of 0.008210 on the test-B-eval set. Please note that these results are obtained after incorporating the test-A-eval set into the training process. We can observe that our PDDWS system significantly outperforms the official baseline, achieving an absolute improvement of up to 93.69\% and securing first place in the challenge.

\begin{table}[t]
\centering
\caption{The score results of each competition system on the test-B-eval set.}
\vspace{5pt}
\renewcommand{\tabcolsep}{0.2cm}
\renewcommand\arraystretch{1.3}
\begin{tabular}{lccc}
\toprule     
\textbf{System}           & \textbf{Score $\downarrow$}            & \textbf{FAR $\downarrow$}               & \textbf{FRR $\downarrow$} \\ 
\midrule 
\textbf{Proposed (Rank 1st)} & \textbf{0.008210}     & \textbf{0.003210}         & \textbf{0.005000}  \\
Rank 2nd Team  &0.009801          & 0.004801 & 0.005000  \\
Rank 3rd Team & 0.010533         & 0.003033 & 0.007500 \\
Rank 4th Team & 0.099282          & 0.020282 & 0.079000 \\
Rank 5th Team & 0.112711          & 0.038878 & 0.073833 \\ \midrule
Official Baseline & 0.130306          & 0.028639 & 0.101667   \\
\bottomrule    
\end{tabular}
\label{sb:expcop}
\end{table}

\subsection{Ablation Study}
\label{ssec:AblationStudy}
In order to verify the effectiveness of the components of our system, we performed ablation experiments on each component. In our experiments on the test-A-eval set, we leverage the availability of ground truth labels to traverse all possible thresholds and identify the optimal values that minimize FRR and FAR. Unfortunately, we do not have access to ground truth labels for the test-B-eval set. Table~\ref{sb:exp22} presents the results of ablation experiments for the threshold selection on the test-A eval set, based on the 2branch-d2v2 model. The experiment results indicate that as the threshold increases from 55 to 61, FAR gradually increases while FRR gradually decreases. The strategy with a threshold rank of 55 achieves the lowest FAR but significantly reduces FRR performance. Conversely, the strategy with a threshold rank of 60 achieves the lowest FRR, with no significant compromise in FAR performance (the score for Thresh rank 60 is better than that for Thresh rank 55). Therefore, in subsequent experiments on the test-B-eval dataset, we primarily focus on Thresh rank of 60.

\begin{table}[t]
\centering
\caption{Performance on the test-A-eval set using different threshold rank.}
\vspace{5pt}
\renewcommand{\tabcolsep}{0.3cm}
\renewcommand\arraystretch{1.3}
\begin{tabular}{cccc}
\toprule     
\textbf{Thresh rank}           &\textbf{Score $\downarrow$}            & \textbf{FAR $\downarrow$}               & \textbf{FRR $\downarrow$} \\ 
\midrule 
55                    & 0.0697          & \textbf{0.0022}              & 0.0675 \\
56                    & 0.0575          & 0.0025              & 0.0550 \\
57                    & 0.0454          & 0.0029              & 0.0425 \\
58                    & 0.0384          & 0.0034              & 0.0350 \\
59                    & 0.0339          & 0.0039              & 0.0300 \\
60                    & \textbf{0.0322}          & 0.0047              & \textbf{0.0275} \\
61                    & 0.0331          & 0.0056              & 0.0275 \\
\bottomrule    
\end{tabular}
\label{sb:exp22}
\end{table}

Table~\ref{sb:exp2} shows the ablation experiments conducted on the test-B eval set. In this experiment, we select a thresh rank of \textbf{60}. 
In the first row of the table, our approach involves feeding the audio input directly into the Paraformer model. This model generates an ASR output (ASR result2), which we then evaluate against a predefined wake-up word list. If the ASR result matches any word in the wake-up word list, we assign the corresponding label to the detection; otherwise, we categorize it as a filler. This straightforward method yields a low FAR, indicating few instances where non-wake-up words are incorrectly detected as wake-up words. However, it also results in a relatively high FRR, indicating instances where actual wake-up words are missed or not detected.
To enhance the accuracy and efficiency of wake-up word detection, we integrate the ASR filter module into our system. This module refines the initial detections by further analyzing the ASR output and cross-verifying it with the wake-up word list. By leveraging the ASR filter module, our system achieves optimal performance metrics, effectively reducing both FAR and FRR, thereby enhancing overall system reliability and user experience.

\begin{table}[t]
\centering
\caption{Ablation study of ASR filter on the test-B-eval set.}
\vspace{5pt}
\renewcommand{\tabcolsep}{0.3cm}
\renewcommand\arraystretch{1.25}
\begin{tabular}{lccc}
\toprule     
\textbf{System}           &\textbf{Score $\downarrow$}            & \textbf{FAR $\downarrow$}               & \textbf{FRR $\downarrow$} \\ 
\midrule 
ASR result2                                       & 0.021101          & \textbf{0.001601} & 0.019500 \\
2branch-d2v2                                     & 0.011934          & 0.004434          & 0.007500 \\
~~~+ASR filter                                      & 0.010822          & 0.003322          & 0.007500 \\
\bottomrule    
\end{tabular}
\label{sb:exp2}
\end{table}
To verify the effectiveness of our proposed 2branch-d2v2 module, we perform ablation experiments on test-A-eval-set.
In the first row of Table ~\ref{sb:exp3}, we do not use CTC loss for regularization, i.e. 1branch-d2v2. It is apparent that the effectiveness suffers without using CTC loss, which justifies the integration of ASR modeling to assist KWS modeling.

\begin{table}[t]
\centering
\caption{Ablation study of 2-branch module on the test-A-eval set.}
\vspace{5pt}
\renewcommand{\tabcolsep}{0.4cm}
\renewcommand\arraystretch{1.25}
\begin{tabular}{cccc}
\toprule     
\textbf{System}           & \textbf{Score $\downarrow$}            & \textbf{FAR $\downarrow$}               & \textbf{FRR $\downarrow$} \\ 
\midrule 
1branch-d2v2                                     & 0.0396          & 0.0046          & 0.0350 \\
2branch-d2v2                                     & \textbf{0.0343}          & \textbf{0.0043}          & \textbf{0.0300} \\
\bottomrule    
\end{tabular}
\label{sb:exp3}
\end{table}

Futhermore, we also perform experiments with different data finetune paraformer on test-A eval set, as shown in Table~\ref{sb:exp4}. Using the same matching strategy as in Table~\ref{sb:exp2}, we can see that finetuning using synthetic data can make the model more adaptable to the dysarthric environment.

\begin{table}[t]
\centering
\caption{Ablation study of Finetuned Paraformer module on the test-A-eval set.}
\vspace{5pt}
\renewcommand{\tabcolsep}{0.3cm}
\renewcommand\arraystretch{1.25}
\begin{tabular}{lccc}
\toprule     
\textbf{System}           & \textbf{Score $\downarrow$}            & \textbf{FAR $\downarrow$}               & \textbf{FRR $\downarrow$} \\ 
\midrule 
paraformer                                     & 0.1050          & 0.0025          & 0.1025 \\
~~+LRDWWS training set                                     & 0.0646          & 0.0021          & 0.0625 \\
~~~~+synthetic data                                      & \textbf{0.0493}          & \textbf{0.0018}          & \textbf{0.0475} \\
\bottomrule    
\end{tabular}
\label{sb:exp4}
\end{table}

\section{CONCLUSION}
\label{sec:con}
In this paper, we introduce our 
Pretrain-based Dual-filter Dysarthria Wake-up word Spotting (PD-DWS) system developed by for the 2024 Low-Resource Dysarthria Wake-Up Word Spotting Challenge. Our system improves performance from two perspectives: audio modeling and dual-filter strategy. We also use TTS data for data augmentation in finetuned Parafomfer module. Finally our system achieves an FAR of 0.00321 and an FRR of 0.005 in the evaluation set of this challenge, ranking first in the competition.

\bibliographystyle{IEEEbib}
\bibliography{strings,refs}

\end{document}